\documentclass[12pt]{article}

\title{Scaling law of Wolff cluster surface energy}
\author{Pai-Yi Hsiao{\small $^{(1)}$} and Pascal Monceau{\small $^{(1,2)}$}}
\begin{document}
\maketitle

\begin{center}
{\small $^{1}$Laboratoire de Physique Th\'{e}orique de la Mati\`{e}re
Condens\'{e}e,\\
P\^ole Mati\`ere et Syst\`eme Complexes (FR2438 CNRS), 
Universit\'{e} Paris 7 -- Denis~Diderot,\\
Case 7020, 2 Place Jussieu, 75251 Paris Cedex 05, France\\
$^{2}$D\'{e}partement de Physique et Mod\'{e}lisation, 
Universit\'{e} d'Evry-Val d'Essonne,\\
Boulevard Fran\c{c}ois Mitterrand, 91025 Evry Cedex, France }
\end{center}

\begin{abstract}
We study the scaling properties of the clusters grown by the Wolff algorithm
on seven different Sierpinski-type fractals of Hausdorff dimension
$1 < d_f \le 3$ in the framework of the Ising model.
The mean absolute value of the surface energy of Wolff 
cluster follows a power law with respect to the lattice size. 
Moreover, we investigate the probability density distribution of 
the surface energy of Wolff cluster and 
are able to establish a new scaling relation. 
It enables us to introduce a new exponent  
associated to the surface energy of Wolff cluster.
Finally, this new exponent is linked to a dynamical exponent
via an inequality. 
 
\end{abstract}
\noindent \textbf{PACS:}\newline
64.60.Ak Renormalization-group, fractal, and percolation studies 
         of phase transitions\\
68.35.Md Surface thermodynamics, surface energies\\
75.40.Mg Numerical simulation studies\\
89.75.Da Systems obeying scaling laws\\

In 1966, Kadanoff proposed an intuitive renormalization picture
\cite{Kadanoff66} to explain the Widom's phenomenological hypothesis 
\cite{Widom65}. 
This hypothesis states a homogeneous transformation of the singular part 
of the free energy per spin  under a change of length unit from 1 to $b$ 
in the vicinity of the critical temperature $T_c$
and can be written as 
\begin{equation}
f(t,h)=b^{-d} f(b^{y_t}t, b^{y_h}h),
\label{Widom_hypothesis}
\end{equation}
where $d$ is the integer space dimension of a translationally invariant system;
the reduced temperature $t=T/T_c -1$ and the external field $h$ 
are supposed to be small; 
$y_t$ and $y_h$ are two exponents associated with the two relevant directions 
of the renormalization flows. 
One notices that a translationally invariant system of dimension $d$ 
is auto-similar and can be considered as a particular case of fractal. 
In the case of a \textit{general} fractal system,
the translational symmetry is broken;
however, a generalization of the space dimension can be done 
by introducing the Hausdorff dimension $d_f$. 
The number of spins on such system can be written as $N=L^{d_f}$
where $L$ is the network size. 
Thus, the Widom's hypothesis can be generalized by 
replacing the factor $b^{-d}$ in Eq.(\ref{Widom_hypothesis}) by $b^{-d_f}$
to describe the decrease of the effective number of spins 
during the change of the length scale.
Since fractals are constructed by iteration of a generating cell, 
$b$ cannot take any value; 
it must be chosen so that the fractal structure remains invariant under 
the change of length unit. 
The critical behaviors of spin models on fractals whose Hausdorff dimension 
lies between $1$ and $3$ have been studied intensively by 
Monceau and Hsiao and co-workers \cite{Monceau98, Hsiao00, Monceau01, Hsiao02} 
and Carmona \textit{et al.} \cite{Carmona98}, 
by performing Monte Carlo simulations. 
The results show the validity of the above generalization.
More recently, a direct verification of the Kadanoff's real space 
renormalization group picture of the Ising model 
in the case of a Sierpinski fractal has been achieved by 
Hsiao and Monceau \cite{Hsiao02_MCRG}; they used 
a Monte Carlo renormalization group method.

The simulation works mentioned above have been mainly performed with 
the help of the Monte Carlo Wolff algorithm \cite{Wolff89a}.
This algorithm is able to reduce efficiently the critical slowing down. 
Instead of involving a single-spin flip at each update of the spin state 
configuration, the Wolff algorithm grows firstly a cluster 
(so-called Wolff cluster) and then updates the state configuration 
by assigning a new state to the spins of this cluster.
Much information on the critical behavior of a discrete spin system
can be brought out from the geometrical properties of the Wolff clusters. 
It has been shown that the mean size of Wolff clusters for the Ising model 
scales as $L^{\gamma/\nu}$ at $T_c$ where $\gamma$ and $\nu$ are
the critical exponents associated to the susceptibility and the correlation
length \cite{Wolff89a}. 
It provides an alternative way to calculate $\gamma/\nu$. 
Monceau and Hsiao checked that the values of 
$\gamma/\nu$ calculated from the Wolff mean cluster sizes are consistent with 
the ones calculated from the behavior of the maxima of the susceptibility 
with respect to $L$ in the case of the Sierpinski fractals 
with Hausdorff dimensions betweeen $1$ and $3$.\cite{Monceau02,Monceau03} 
Moreover, the most striking result they carried out is 
the scaling invariance of the Wolff cluster size probability densities 
${\cal P}(s)$ under a suitable rescaling. 
This scaling law holds at the critical point and reads
\begin{equation}
{\cal P}(s) = b^{-d_f} {\cal P}(b^{-x} s), 
\label{Scaling_prob_s}
\end{equation}
where $s$ denotes the size of Wolff cluster associated with a Monte Carlo step 
and $b$ is some appropriate change of length scale which keeps invariant 
the fractal structure. 
They proved that the exponent $x$ is equal to $y_h=\beta/\nu+\gamma/\nu$
where $\beta$ is the critical exponent associated to the magnetic moment. 
This scaling law can be understood as a decrease of Wolff cluster size 
by a factor $b^{y_h}$ under the change of the length scale $1 \to b$. 
During the simulations of the Ising model, 
the total magnetization $M$ and the total energy $E$ are calculated
at each update of the spin state configuration.  
As a matter of fact, the absolute difference $|\Delta M|$ between 
two successive values of $M$ is equal to 
two times the size of the Wolff cluster grown during a Monte Carlo step.
Hence, the scaling law described in Eq.~(\ref{Scaling_prob_s}) is satisfied by 
the probability density distribution of $|\Delta M|$. 
It is worth noticing that the mean absolute difference 
$|\Delta E|$ between two successive values of the total energy 
represents two times the surface energy of the Wolff cluster. 
Thus, there is much to learn in studying the scaling properties of $|\Delta E|$ 
and the associated probability density distribution. 
The purpose of this paper is to study the behavior of the surface energy of 
Wolff clusters grown on fractal lattices at the critical point.  
We investigate seven different Sierpinski-type fractals of Hausdorff dimension 
lying between 1 and 3. 
The lattices are generated iteratively from some generating cells 
and denoted $SP(\ell^d, n_{occ}, k)$. 
$\ell$ is the size of the generating cell, 
$d$ is the integer space dimension in which the lattice is embedded, 
$n_{occ}$ is the number of occupied sites in the generating cell, 
and $k$ is the number of iteration steps. 
Ising spins are placed at each center of the occupied sites. 
The size of $SP(\ell^d, n_{occ}, k)$ is $L=\ell^k$ and 
the number of spins is $N=n_{occ}^k$. 
The Hausdorff dimension is defined by 
$d_f=\log N/\log L=\log n_{occ}/\log \ell$. 
The \textit{true} mathematical fractal is obtained 
only when $k$ tends to infinity; 
in this case, we denote it by $SP(\ell^d, n_{occ})$. 
The structure of the fractal is not indicated in these symbols. 
The seven generating cells $SP(\ell^d, n_{occ}, 1)$ we have chosen are 
\begin{enumerate}
\item[(1).] $SP(2^2, 4, 1)$: a 2 by 2 square, 
\item[(2).] $SP(3^2, 8, 1)$: a 3 by 3 square where the center sub-square is
            removed, 
\item[(3).] $SP(5^2, 24, 1)$: a 5 by 5 square where the center sub-square is
            removed, 
\item[(4).] $SP(2^3, 8, 1)$: a cube of size 2, 
\item[(5).] $SP(3^3, 26, 1)$: a cube of size 3 where the center sub-cube 
            is removed,
\item[(6).] $SP(4^3, 56, 1)$: a cube of size 4 where the center sub-cubes of
            size 2 are removed,
\item[(7).] $SP(3^3, 18, 1)$: one removes in addition the 8 sub-cubes
            at the corners of $SP(3^3, 26, 1)$.
\end{enumerate}
$SP(2^2,4)$ and $SP(2^3, 8)$ are exactly a square and a cubic lattice 
of infinite size. 
It has been shown that the Ising model exhibits a second order ferromagnetic 
phase transition on a fractal, provided that the lattice 
has a particular geometrical property:
the ramification order must be infinite \cite{Gefen84}.
It is the case for the fractals studied here. 
Moreover, the critical temperatures $T_c$ on these fractals are available 
\cite{Hsiao00,Monceau01,Binder01,Baxter82} and 
their values are recalled in the table 1.
The mean absolute value of the Wolff cluster surface energy  
can be calculated from the relation
\begin{equation}
\frac{1}{2} \langle |\Delta E| \rangle = 
\frac{1}{2 (N_{sim}-1)} \sum_{n=1}^{N_{sim}-1} |E_{n+1}-E_n|  
\label{EMEAN}
\end{equation}
where $N_{sim}$ is the total number of Monte Carlo steps 
and $E_n$ is the total energy 
of the \textit{n-th} updated configuration.
Firstly, we found that $\langle |\Delta E| \rangle$ follows 
power laws at $T_c$ with respect to the lattice size $L$ 
in the case of the seven different fractals we investigated. 
It enables to define a surface exponent:  
$\langle |\Delta E| \rangle \sim L^{S_W}$.  
Fig.1 shows the behavior of $\langle |\Delta E| \rangle$ as 
a function of $L$ in logarithmic coordinates. 
The points line up along straight lines except for the small sizes, 
where the scaling corrections due to the finite-size effects are expected. 
It has been suggested that these corrections have a topological 
character and are linked to the slow convergence towards the thermodynamical 
limit in the case of the fractals with broken translational symmetry 
\cite{Monceau01}.  
As a matter of fact, $\langle |\Delta E| \rangle$ follows perfectly
a power law in the case of the translationally invariant lattices 
$SP(2^2,4,k)$ and $SP(2^3,8,k)$  
where $L$ increasing as a geometrical series  covers many orders of magnitude. 
We report the measured surface exponent $S_W$ in the table 1, 
where least-square fits are performed
from $k=8$ to $12$ for $SP(2^2, 4,k)$, 
from $k=5$ to  $8$ for $SP(2^3, 8,k)$, 
from $k=4$ to  $8$ for $SP(3^2, 8,k)$, 
from $k=2$ to  $4$ for $SP(4^3,56,k)$,  
from $k=3$ to  $5$ for $SP(5^2,24,k)$, for $SP(3^3,18,k)$, and 
for $SP(3^3,26,k)$, respectively. 
On the other hand, since Monte Carlo simulations can be performed only 
on lattices of finite size and we have omitted the small-size data 
when extrapolating the thermodynamical limit, 
a slow crossover behavior may be interpreted as an asymptotic one. 
A detailed study shows that the points in Fig.1 exhibit a very slight 
concavity.  
The reported value $S_W$, hence, should be taken as an upper bound 
in a strict sense; 
however, the \textit{real} thermodynamical-limit value is expected to 
be very close to the reported one. 

We are now able to go further and study the scaling properties of 
the probability density distributions ${\cal P}(|\Delta E|)$ of 
Wolff cluster surface energy.
The curves showing ${\cal P}(|\Delta E|)$ are similar to each other
in logarithmic coordinates. 
These results suggest to write down a homogeneous transformation 
under the form: 
\begin{equation}
{\cal P}(|\Delta E|) = b^{-D_S} {\cal P}(b^{-y_S} |\Delta E|),
\label{Scaling_prob_dE}
\end{equation}   
where $D_S$ and $y_S$ are some introduced exponents. 
An intuitive guess is to set $D_S$ equal to $d_f-1$
since it describes the usual surface dimension of a system of
bulk dimension $d_f$.
According to Eq.(\ref{Scaling_prob_dE}), $\langle |\Delta E| \rangle$ should 
scale as :
\begin{eqnarray}
\nonumber
\langle |\Delta E| \rangle 
&=& \int_0^{\infty} |\Delta E| {\cal P}(|\Delta E|)\, d|\Delta E|\\
\nonumber
&=& \int_0^{\infty} |\Delta E| b^{-(d_f-1)}{\cal P}(b^{-y_S}|\Delta E|) \, 
d|\Delta E|\\ 
\nonumber
&=& b^{-(d_f-1)+2y_S} \int_0^{\infty} |\Delta E'|  {\cal P}(|\Delta E'|)\, 
d|\Delta E'|, 
\end{eqnarray}
where we have performed the change of variable $|\Delta E'|=b^{-y_S}|\Delta E|$.
Since the correlation length is divergent at $T_c$, 
we can set $b$ equal to the lattice size $L=\ell^k$. 
$\langle |\Delta E| \rangle$ is, therefore, proportional to 
$L^{-(d_f-1)+2y_S}$. 
We are, hence, able to obtain a relation linking 
the exponents $S_W$ and $y_S$:    
\begin{equation}
S_W = -(d_f-1) + 2 y_S.
\label{S_W_y_S}
\end{equation}
The values of $y_S$ calculated from Eq.(\ref{S_W_y_S}) for the seven fractals 
investigated are given in the table 1.

The similarity property of the curves showing the probability density 
distributions ${\cal P}(|\Delta E|)$ and 
the validity of the homogeneous transformation Eq.(\ref{Scaling_prob_dE}) 
can be brought out in the following way: 
for a given structure at different values of the iteration step $k$,  
the curves showing ${\cal P}(|\Delta E|)$ collapse onto the one 
corresponding to the lattice of the largest size $L=\ell^{k_{max}}$ 
under the mapping: 
\begin{eqnarray}
\nonumber
(|\Delta E|\ ,\ {\cal P}(|\Delta E|)) &\to& 
(\ell^{(k_{max}-k)y_S} |\Delta E|\ ,\  
\ell^{-(d_f-1)(k_{max}-k)} {\cal P}(|\Delta E|))\ . 
\end{eqnarray} 
Fig.2 shows such collapses for the three largest values of $k$ 
on the fractals $SP(3^2,8)$, $SP(2^2,4)$, and $SP(3^3,18)$. 
These data-collapses work out in a reliable way with the 
values of $y_S$ given in Table 1.  
It confirms that $D_S$ is equal to $d_f-1$
(It has also been checked for the four other fractals). 
Furthermore, ${\cal P}(|\Delta E|)$ does not exhibit a peak as 
${\cal P}(s)$ does. 
The effect of segregation between large and small clusters mentioned 
in Ref.\cite{Monceau02,Monceau03} is smoothed in the behavior of 
${\cal P}(|\Delta E|)$. 
It means that an important part of the simulation is carried out in updating 
large clusters;  
however, the update of the large clusters does not necessarily imply a 
large change of the total energy. 
For a given cluster size, the probability distribution of the border surface 
is broad. 

A fundamental question is to know if the introduced exponent $y_S$ 
is a new one, that is, if it is independent of the two renormalization 
group eigen-exponents $y_t$, $y_h$ and the Hausdorff dimension $d_f$. 
To our knowledge, it's the first time that the surface energy scaling
property of Wolff cluster is investigated. 
No analytical or theoretical treatment is available. 
According to our Monte Carlo simulation results (table 1),
$y_S$ seems to be independent of $y_t$, $y_h$ and $d_f$. 
Whether $y_S$ links to some of the surface exponents of a bulk system
\cite{Binder83_Diehl86} remains an open question.
In the table 1, regardless of the fractal structures, 
one can find that the values of $y_t$, $y_h$ and $y_S$ increase 
as the Hausdorff dimension $d_f$ increases. 
Moreover, we do not expect that $\epsilon$-expansion results can be 
interpolated to non-integer dimensions and  provide the values 
of $y_t$ and $y_h$ on the fractals, even the value of $y_S$. 
Accurate Monte Carlo studies have shown that the universality of phase
transitions on the hierarchical lattices without translational symmetry
should depend on the lattice structure \cite{Hsiao00, Monceau01}. 
Hence, $y_S$ associated to the fractals with broken translational symmetry
depends on the lattice structure too. 

As the Wolff clusters are dynamical objects, we exploit here the connection 
between the surface exponent $y_S$ and the dynamical scaling exponent.
We firstly study the mean square surface energy of Wolff cluster 
\begin{equation}
\langle (\Delta E)^2 \rangle = 
\frac{1}{N_{sim}-1} \sum_{n=1}^{N_{sim}-1} (E_{n+1}-E_n)^2 \ .
\end{equation}
We find that $\langle (\Delta E)^2 \rangle$ 
lines up along straight lines with respect to $L$  
except for the small lattice sizes  
in logarithmic coordinates for the seven fractals investigated (see Fig.3).
The associated exponent $u$, defined by 
$\langle (\Delta E)^2 \rangle \sim L^{2u}$, could be measured
in the same way as described in the previous text for the exponent $S_W$;
the values of $u$ are reported in the table 1.
We, then, express $\langle (\Delta E)^2 \rangle$ as 
$2\left[\langle E^2 \rangle-\langle E\rangle^2 \right]
\left(1-\theta_{EE}(1)\right)$
where $\theta_{EE}(n)=(\langle E_{0}E_{n} \rangle-\langle E\rangle^2)/
(\langle E^2 \rangle-\langle E\rangle^2)$ is the normalized autocorrelation 
function.
Since the autocorrelation time is much longer than 1
(in some appropriate time unit),
$1-\theta_{EE}(1)$ represents approximatively the negative derivative of the  
autocorrelation function at the origin.
It enables to define a statistic-fluctuation autocorrelation time 
$\tau_{sf}^E$  and an associated dynamical exponent $z_{sf}^E$: 
$-(d\theta_{EE}(t)/dt|_{t=0}) = (\tau_{sf}^E)^{-1} 
\sim L^{-z_{sf}^E}$.\cite{Landau00}
Notice that, according to Eq.(\ref{Widom_hypothesis}),
the term $\langle E^2 \rangle-\langle E\rangle^2$ should asymptotically scale 
as $L^{\alpha/\nu+d_f}=L^{2/\nu}$ at $T_c$, where $\alpha$ is the critical 
exponent associated to the specific heat per spin $c_v$. 
One should keep in mind that the non-singular part of the free energy per spin
gives, in addition, an important contribution in determination of
the critical behavior of $c_v$ \cite{Hsiao00}. 
We, therefore, have $2u=2\nu^{-1}-z_{sf}^E$ in the thermodynamical limit.
With the help of the inequality 
$\langle |\Delta E| \rangle  \le \sqrt{\langle (\Delta E)^2 \rangle}$,
we can link the surface exponent $S_W$ with the dynamical one $z_{sf}^E$
in the following way: $S_W$ is upper bounded by $u=\nu^{-1}-z_{sf}^E/2$. 
Moreover, since $\langle (\Delta E)^2 \rangle$ follows a power law,
a homogeneous transformation for the probability density of $(\Delta E)^2 $ 
can be stated;  
it takes the following form
\begin{equation}
{\cal P}((\Delta E)^2) = b^{-2(d_f-1)} {\cal P}(b^{-y} (\Delta E)^2),
\end{equation}   
and has been verified on the seven fractals investigated here.
In this case, $(\Delta E)^2$ decreases by a factor $b^y=b^{u+(d_f-1)}$ under
a suitable change of length unit $1 \to b$.  
A similar process could be applied in the study of Wolff cluster size
or the successive change of total magnetic moments. 
One can show that $\langle |\Delta M| \rangle  \le \sqrt{\langle (\Delta M)^2
\rangle}$ and, therefore, $\gamma/\nu$ is upper bounded by 
$(\gamma/\nu+d_f-z^M_{sf})/2$ where $z^M_{sf}$ is the 
statistic-fluctuation dynamical exponent associated to 
the total magnetic moment. 
It implies  $z^M_{sf} \le d_f-\gamma/\nu =2\beta/\nu$. 

The method developed in this paper can be generalized to the study of some 
physical quantity whose mean value follows a power law with respect to $L$. 
For instance, since $\langle |M| \rangle \sim L^{y_h}$ at $T_c$, 
one can verify, by constructing a homogeneous transformation for
the probability density of $|M|$ similar to Eq.(\ref{Scaling_prob_s}),
that $|M|$ decreases by a factor $b^{(y_h+d_f)/2}$ 
under a suitable rescaling $1 \to b$. 
All these results suggest that the scaling properties of 
some physical quantity at $T_c$, where the correlation length is divergent, 
comes originally from the underlying hierarchal structure of a fractal. 

In summary, we have studied the scaling properties of the Wolff cluster 
surface energy in the framework of the Ising model in the case of seven 
different fractals dimensions. 
A new scaling relation for the absolute value of the surface energy 
$|\Delta E|$ of the Wolff's cluster has been established. 
We have shown that $|\Delta E|$ scales as $b^{-y_S}|\Delta E|$ under
an increment of the length unit by the factor $b$, 
which remains invariant the underlying structure of the fractal. 
Finally, the surface exponent $S_W$ is proved to be upper bounded by
$\nu^{-1}-z^E_{sf}/2$.

%%--------- Acknowledgements ---------------
\noindent
\textbf{Acknowledgements}
A part of the numerical simulations was carried out in the 
\textit{Institut de D\'eveloppement et des Ressources en Informatique 
Scientifique}, supported by the 
\textit{Centre National de la Recherche Scientifique} (project No.~$021186$). 
We acknowledge the scientific committee and the staff of the center. 
We are also grateful to the \textit{Centre de Calcul Recherche} 
of the University Paris VII---Denis Diderot, 
where the rest of the simulations have been performed.

 %------------------------------------------------------

\clearpage %-----------------------------------------------------------------
\section*{Table}
\begin{table}[htpb]
\caption{
Measured values of $S_W$ and $u$ on the seven fractals.
$y_S$ is calculated from Eq.(\ref{S_W_y_S}).
The values of $T_c$, $y_t$ and $y_h$ are recalled from 
Ref.\cite{Hsiao02_MCRG}, Ref.\cite{Monceau01}, 
Ref.\cite{Baxter82}, Ref.\cite{Hsiao00}, and Ref.\cite{Binder01}.  
}
\begin{tabular}{|c||c|c|c||c|c||c|}
\hline
fractal &  $T_c$ & $y_t$ & $y_h$ & $S_W$ & $y_S$  & $u$ \\ \hline
$SP(3^2, 8)$ & 1.4795(5) & 0.449(6)   & 1.8198(11) & 0.838(2)  & 0.865(1) & 0.878(4)\\ 
$SP(5^2,24)$ & 2.0660(15) & 0.923(2)  & 1.861(5)   & 0.8188(2) & 0.8967(1)& 0.880(2)\\ 
$SP(2^2, 4)$ & $2/\ln(1+\sqrt{2})$ & 1  & 1.875      & 0.8100(2) & 0.9050(1) & 0.8777(8)\\ 
$SP(3^3,18)$ & 2.35090(9) & 1.185(16) & 2.317(16)  & 0.849(8)  & 1.240(4) & 1.016(9)\\ 
$SP(4^3,56)$ & 3.99893(10)& 1.410(36) & 2.407(14)  & 0.742(10) & 1.323(5) &0.990(11)\\ 
$SP(3^3,26)$ & 4.21701(6)	& 1.503(53) & 2.449(22)  & 0.738(2)  & 1.352(1)& 0.995(5)\\ 
$SP(2^3,8)$  & 4.511516(41)& 1.588(2) & 2.482(2)  & 0.759(5)  & 1.3795(25)& 1.018(5)\\
\hline
\end{tabular}
\end{table}                     

\clearpage %----------------------------------------------------------------- 
\section*{Figure captions}
\begin{enumerate}
\item[Fig.1] 
$\langle |\Delta E|\rangle$ versus lattice size $L$ 
on the seven fractals investigated in logarithmic coordinates. 
\item[Fig.2]
Collapses of ${\cal P}(|\Delta E|)$ on the fractals 
$SP(3^2,8,k)$, $SP(2^2,4,k)$ and $SP(3^3,18,k)$ under the mapping 
described in the text.
\item[Fig.3] 
$\sqrt{\langle (\Delta E)^2\rangle}$ versus lattice size $L$ 
on the seven fractals investigated in logarithmic coordinates. 
\end{enumerate}

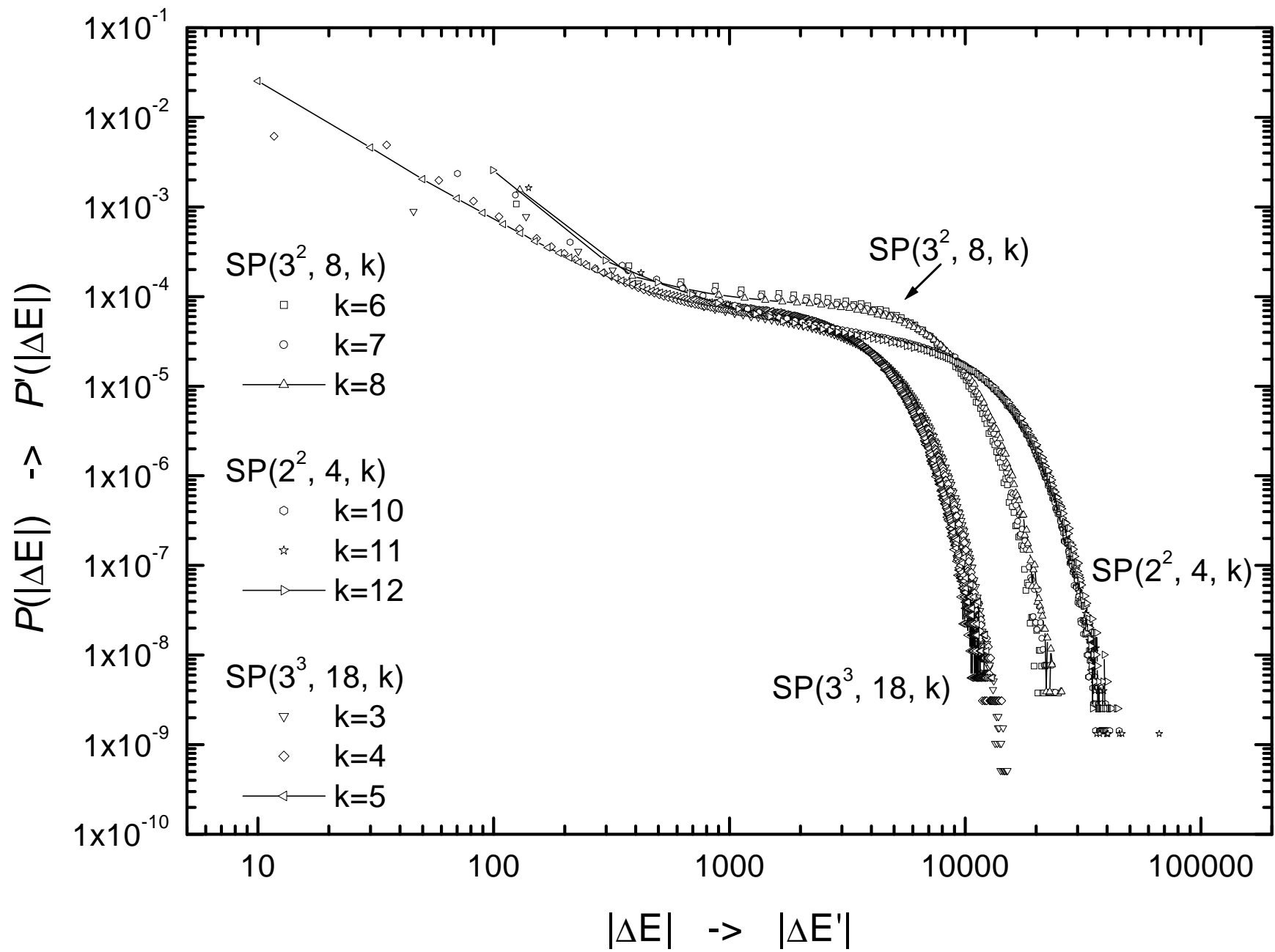
\end{document}